**Article**

# Early systems change necessary for catalyzing long-term sustainability in a post-2030 agenda

## Graphical abstract

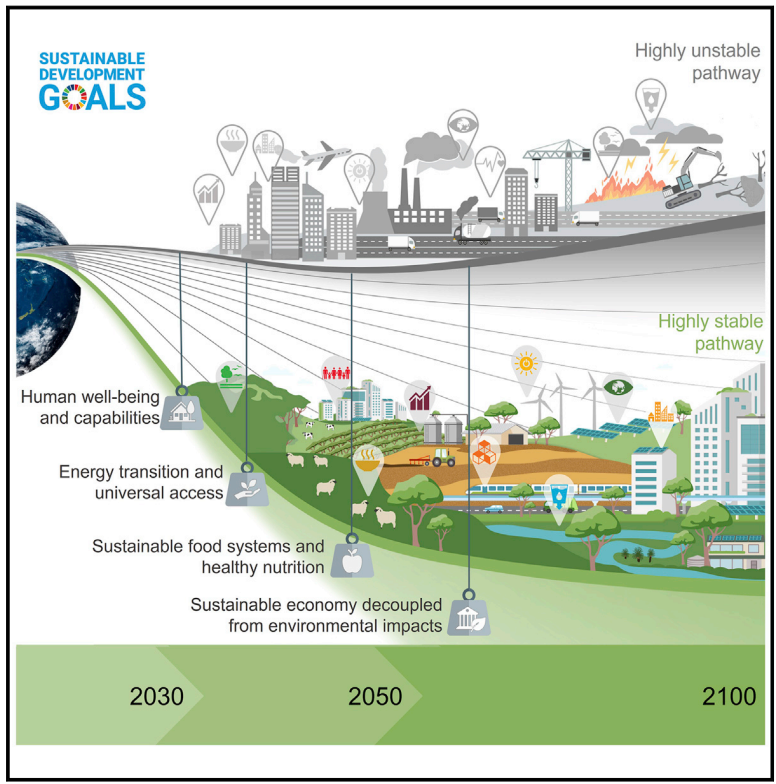

## Authors

Enayat A. Moallemi, Sibel Eker, Lei Gao, ..., Michael Obersteiner, Zhaoxia Guo, Brett A. Bryan

## Correspondence

e.moallemi@deakin.edu.au

## In brief

Our model-based projection of progress toward the Sustainable Development Goals shows a limited chance of near-term success for achieving the 2030 Agenda. However, complex systemic interactions and feedbacks mean that early planning and action towards systems change can accelerate progress towards even more ambitious 2050 and 2100 targets than those for 2030. This longer-term analysis is important to improve the understanding of the conditions that appear to make a limited contribution to initial progress by the 2030 milestone but can become increasingly influential later in the century.

## Highlights

- Sustainability progress to 2100 was analyzed using a system dynamics model

- Progress towards 2030 targets was limited across most pathways

- Early intervention is important for accelerating progress towards increasingly ambitious sustainability targets by 2050 and 2100

*CellPress*

## Article

# Early systems change necessary for catalyzing long-term sustainability in a post-2030 agenda

Enayat A. Moallemi,[1,10,*] Sibel Eker,[2,3] Lei Gao,[4] Michalis Hadjikakou,[1] Qi Liu,[3,5] Jan Kwakkel,[6] Patrick M. Reed,[7] Michael Obersteiner,[8] Zhaoxia Guo,[5,9] and Brett A. Bryan[1]

[1]Centre for Integrative Ecology, School of Life and Environmental Sciences, Deakin University, Melbourne, VIC, Australia
[2]Nijmegen School of Management, Radboud University, Nijmegen, the Netherlands
[3]International Institute for Applied Systems Analysis, Laxenburg, Austria
[4]The Commonwealth Scientific and Industrial Research Organisation (CSIRO), Waite Campus, Urrbrae, South Australia, Australia
[5]Business School, Sichuan University, Chengdu 610065, P.R. China
[6]Faculty of Technology, Policy and Management, Delft University of Technology, Delft, the Netherlands
[7]Department of Civil and Environmental Engineering, Cornell University, Ithaca, NY, USA
[8]The Environmental Change Institute, University of Oxford, Oxford, UK
[9]Soft Science Institute, Sichuan University, Chengdu 610065, P.R. China
[10]Lead contact
*Correspondence: e.moallemi@deakin.edu.au
https://doi.org/10.1016/j.oneear.2022.06.003

**SCIENCE FOR SOCIETY** The world is subject to multiple global challenges, including climate change, environmental degradation, poverty, and inequality. The United Nations Sustainable Development Goals (SDGs) represent nations' collective ambition to overcome these challenges and achieve a more prosperous and sustainable future for all by 2030. However, with less than 8 years remaining, assessments have concluded that it is unlikely that the SDGs will be fully achieved by the end of the decade and that the slowing and reversal of negative trends in key challenges such as climate change is not likely to happen until after 2030. Despite long-term analyses of these component challenges, a deeper and more integrated understanding of the available opportunities to accelerate and achieve sustainability throughout the 21st century is now urgent. This new study, based on the simulated futures of the SDGs, characterizes the scale and feasibility of necessary systems change and provides a guide for long-term progress in sustainability.

## SUMMARY

Progress to date toward the Sustainable Development Goals (SDGs) has fallen short of expectations and is unlikely to fully meet 2030 targets. Past assessments have mostly focused on short- and medium-term evaluations, thus limiting the ability to explore the longer-term effects of systemic interactions with time lags and delay. Here we undertake global systems modeling with a longer-term view than previous assessments in order to explore the drivers of sustainability progress and how they could play out by 2030, 2050, and 2100 under different development pathways and quantitative targets. We find that early planning for systems change to shift from business as usual to more sustainable pathways is important for accelerating progress toward increasingly ambitious targets by 2030, 2050, and 2100. These findings indicate the importance of adopting longer-term timeframes and pathways to ensure that the necessary pre-conditions are in place for sustainability beyond the current 2030 Agenda.

## INTRODUCTION

The United Nations 2030 Agenda (also known as Sustainable Development Goals – SDGs) provides a framework for human development within planetary boundaries through a complementary set of goals (i.e., broad ambitions), targets (i.e., specific thresholds defining success), and indicators (i.e., metrics by which progress toward targets can be judged).[1] Progress to date toward the SDGs has been limited.[2,3] With less than 8 years to go, the scientific community has taken significant steps toward understanding and planning for the SDGs through different approaches, such as future pathway modeling,[4–7] science-based target setting,[8] and SDG interaction analysis.[9–11] SDGs have also been studied at global,[6,12] national,[13,14] and local scales.[10,15] Despite important efforts, past SDG assessments have remained focused mostly on short-term (i.e., 2030) and medium-term (i.e., 2050) evaluations. Many of these assessments have found that (even the most ambitious) pathways are unlikely to fully achieve all SDGs by 2030[4,7,14] and, in some cases, not even by 2050.[12] Although such short/medium-term assessments can be justified in some cases (e.g., for SDGs related to peace, institution, and implementation with significant future uncertainties), they can limit the understanding of longer-term progress and overlook the role of delayed effects and non-linear behavior of slow sustainability trends, driven by systemic feedback interactions that emerge throughout to 2100. This knowledge gap has become increasingly important, given longer-term analyses in neighboring fields (e.g., conservation science,[16] climate change,[17] and demographic studies[18]) and the common finding that the slowing and eventual reversal of negative trends in key sustainability components (e.g., biodiversity loss, greenhouse gas emissions, and population growth) are likely to happen after 2030 and throughout the century.

Here we analyzed long-term global sustainability progress through the SDG lens by 2030, 2050, and 2100 across plausible socioeconomic and environmental development trajectories and through endogenous modeling of inter-connections in human-natural systems. Short-, medium-, and long-term timeframes were aligned with the 2030 Agenda, the Paris Agreement (2050), and the Intergovernmental Panel on Climate Change (IPCC) climate change milestone (2100), respectively. We used this longer-term analysis to show what lies beyond the 2030 Agenda in terms of non-linear progress toward increasingly ambitious targets over time and to identify systems change important for accelerating sustainability progress later in the century. Understanding the required systems change, the opportunities to initiate and sustain it, and the potential barriers to achieving it is prerequisite for future planning to enable missed 2030 targets to be met and exceeded later on and ensure that earlier achievements are not lost through complacency and despair.

## RESULTS

### Global system dynamics modeling for pathway simulation

Our analysis is underpinned by an established model, called Functional Enviro-economic Linkages Integrated Nexus (FeliX). It is developed based on the system dynamics methodology[19] and simulates global-scale social, economic, and environmental interactions and feedback[20] (Figure 1; Table S5; Experimental procedures). FeliX supports modeling of indicators representing eight SDGs related to sustainable food (SDG 2), health and well-being (SDG 3), quality education (SDG 4), clean energy (SDG 7), economic growth (SDG 8), responsible consumption and production (SDG 12), climate action (SDG 13), and life on land (SDG 15). With relatively simple but transparent structure and fast simulation runs, it can cover multiple sustainability dimensions in one integrated modeling framework, which is ideal to support simulation of evolving trade-offs and synergies between human activities (i.e., demography, economy, energy, land, and food) and environmental change (i.e., biodiversity, carbon cycle, and climate systems) over time. Among the few system dynamics models,[12,14,21] FeliX was selected for its transparency[20,22] and credibility in analyzing multiple sustainability dimensions such as emissions pathways,[23] sustainable diet shift,[24] and socio-environmental impacts in human and Earth systems[25] (see Discussion and conclusions for the model's strengths and limitations).

A wide range of long-term development pathways have been assessed, spanning different mitigation policies across systems and with different degrees of compatibility with the Paris Agreement and sustainable development.[7,26] We evaluated a set of five illustrative pathways, in line with the shared socioeconomic pathways (SSPs)[27] and representative concentration pathways (RCPs),[28] as benchmarks for long-term global development trajectories. The selected pathways were not meant to cover all possible futures but to demonstrate the effects of some of the future choices on socioeconomic and environmental development in our analysis. The five pathways are aligned with commonly used SSP-RCP combinations,[26] including Business As Usual (SSP2-4.5; the reference pathway), Green Recovery (SSP1-2.6), Fragmented World (SSP3-7.0), Inequality (SSP4-6.0), and Fossil-Fueled Development (SSP5-8.5) (see Discussion and conclusions for limitations and opportunities of the selected pathways).

We followed a "story and simulation" approach, where the SSP-RCP pathway narratives (Table S1) were used to specify the initial conditions of the model, and then used the model for simulating pathways in quantitative terms (Table S2). Using the FeliX model, we simulated 10,000 model evaluations (called pathway realizations) for each of the five pathways (50,000 pathway realizations in total) to take into account model parameter uncertainty (e.g., natural variability and error in quantification) and explore the variation around the five main pathways for more robust insights (Experimental procedures).

Of the five pathways assessed (Figures 2 and S1), Business As Usual as our reference pathway to 2100 used the continuation of the current trajectories as input assumptions, and therefore its socioeconomic and environmental behavior followed SSP2-4.5 projections. Compared with Business As Usual, Green Recovery had improving socioeconomic trajectories (driven by low population growth, growing economy, and better education access assumptions), fast transition to renewable energy (driven by lower production costs, higher investment, and technology improvement assumptions), and limited land use change (because of lower demand for food, lower meat consumption, and higher agricultural productivity assumptions). The environmental effects of these positive socioeconomic trends together with ambitious climate policies resulted in low deforestation and low-range greenhouse gas (GHG) emissions by 2100. Fragmented World projected declining socioeconomic prosperity (driven by increasing population and slower economic growth), large energy production from fossil fuels, high land use change, and significant environmental footprints (because of high deforestation and emissions levels) compared with Business As Usual. Inequality with slightly better trajectories resulted in moderately improved socioeconomic projections compared with Business As Usual and Fragmented World, relatively slow clean energy transitions, and relatively high food production and land use change trajectories. Fossil-Fueled Development projected an improving socioeconomic future (similar to Green Recovery) but at the cost of unsustainable environmental trajectories

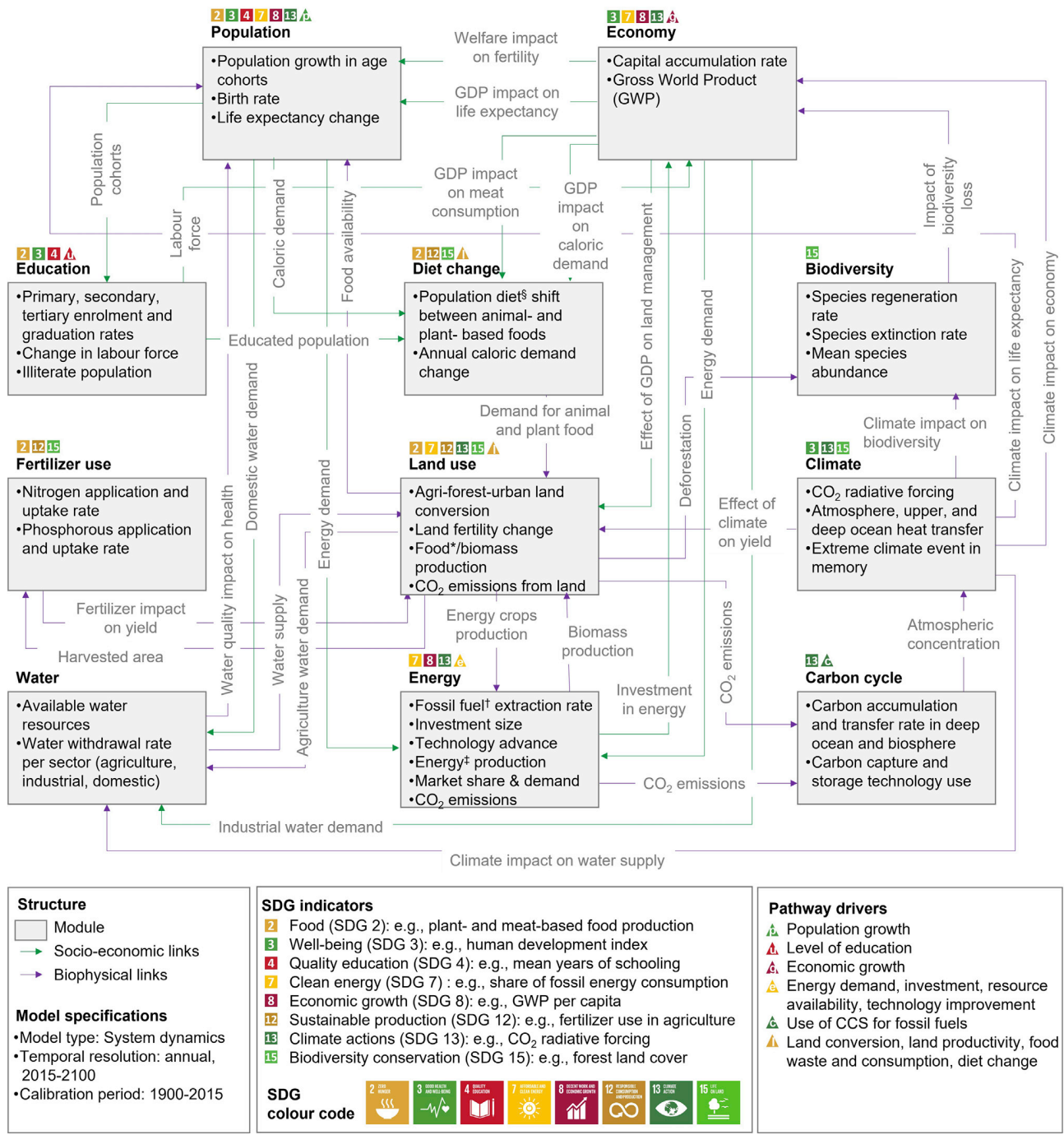

**Figure 1. Overview of the FeliX model**

Gray-shaded boxes represent different sectoral modules in FeliX. Square and triangle markers show where in the model the SDG indicators and pathway drivers were implemented. The marker colors are consistent with their corresponding SDG color, and their annotated numbers/letters correspond to the name of the SDGs and pathway drivers. *Food categories include animal products comprising crop-based meat (poultry and pork), pasture-based meat (beef, sheep, and goat), dairy, and eggs and the supply of plant-based products, including grains, pulses, oil crops, vegetables, roots, and fruits. †Fossil fuels include coal, gas, and oil. ‡Energy includes fossil and renewable (solar, wind, and biomass) energies. §Diet categories include five diet compositions of high to low meat and vegetable consumption. CCS, carbon captured and storage of fossil fuels. See Experimental procedures and Table S5 for more details about the model.

(e.g., slow clean energy transitions and high emissions) because of assumptions of fossil fuel dependency and resource-intensive development.

The results showed consistent behavior with input assumptions across systems in each pathway (Table S1) and also in harmony with the 2100 projections of other integrated assessment models.[29] However, because of the difference in FeliX's model structure and scenario parameter settings, pathways were often quantitatively different (and sometimes with different trajectory patterns) from the outputs of other models. FeliX is structurally different from most integrated assessment models because it is a descriptive model instead of prescribing cost-optimal choices, and it does not assume market equilibrium (see more in Discussion and conclusions). Similar variations in future projection have been observed among other models[29] (see other model projections in Figures 2 and S1), and this highlights the importance of diversifying models to obtain a broader variety of future possibilities for a robust assessment and better appreciation of the deep uncertainty in future projections.[30–32]

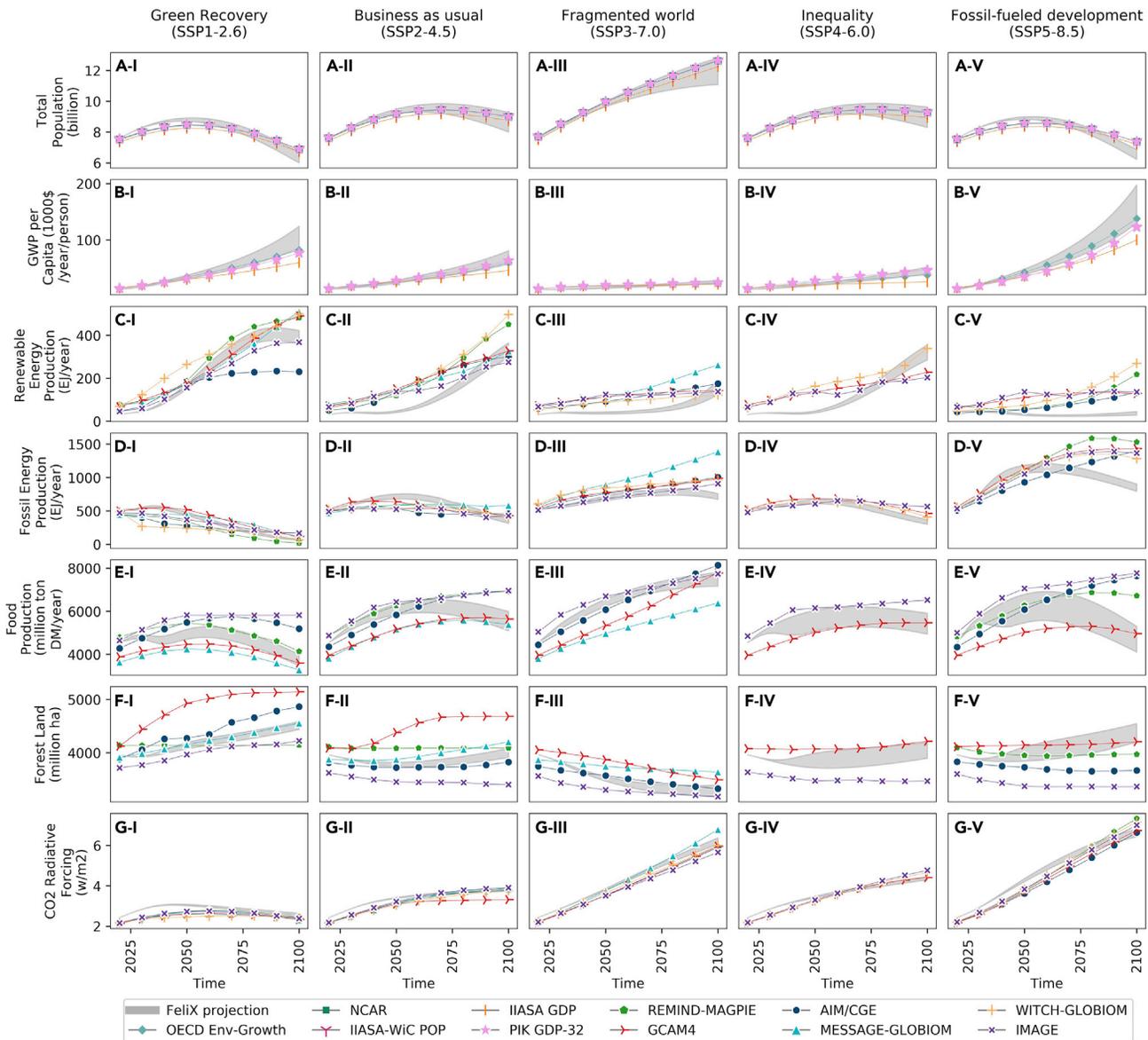

**Figure 2. Pathway simulation results against a suite of seven socioeconomic and environmental model outputs and comparison against similar simulation outputs of major models**
FeliX simulations cover the period 2015–2100 at an annual time step (Experimental procedures). The y axis in all panels represents control variables we used for cross-validating FeliX projections with those of other models. GWP, gross world product.

Our outputs differed from other models mostly in two main areas. First, FeliX projected a faster decline in fossil energy production (e.g., in Fossil-Fueled Development), which resulted from bolder assumptions about fossil fuel and renewable energy production costs. Second, lower livestock production and crop demand were projected in Green Recovery because of FeliX's endogenous diet change assumptions. More explicit assumptions about a shift toward sustainable diets in FeliX, driven by modeling of behavior change and consumption patterns, resulted in lower meat consumption and limited arable land expansion in some pathways compared with outputs from other models (Figures 2 and S1).

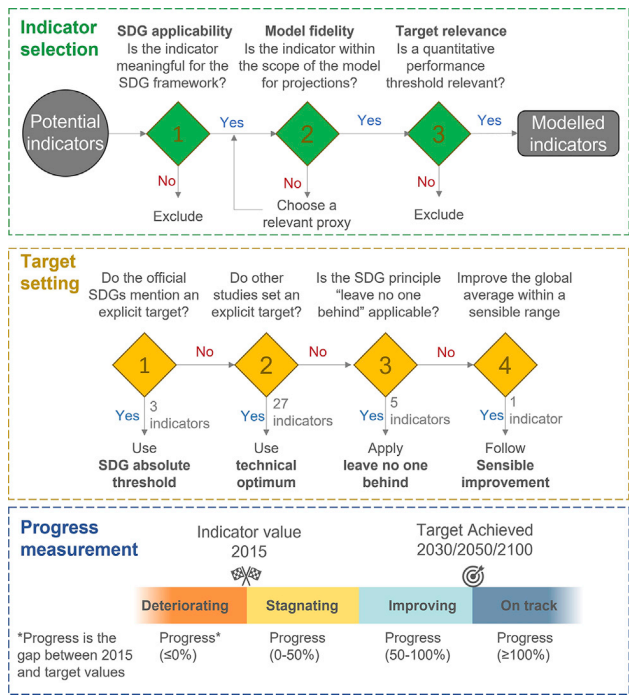

**Figure 3. Indicator selection, target setting, and progress measurement processes**
See Experimental procedures for further details.

## Accelerating SDG progress

We specified a set of well-defined socioeconomic and environmental indicators and targets to measure long-term SDG progress in the projected pathways. Although the current 2030 Agenda has 169 targets and 232 indicators, many are complex, and some lack the specificity to support quantitative projections.[8] We therefore defined 36 complementary sustainability indicators and set quantitative targets related to eight SDGs that were within the scope of our modeling but were also diverse enough to cover most of the key areas of sustainable development related to people (SDGs 3 and 4), prosperity (SDG 8), sustainable resource management (SDGs 2, 7, and 12), and planet integrity (SDGs 13 and 15), as defined by van Vuuren et al.[8] (see Discussion and conclusions for strengths and limitations of selected SDGs).

Indicators were chosen based on a selection process that considers SDG suitability and measurement feasibility within our model (Figure 3; Experimental procedures). We set short-, medium-, and long-term measurable target values with an increasing ambition for 2030, 2050, and 2100 to indicate shifting performance thresholds for the selected indicators over time (Tables S3 and S4). The targets were set based on criteria that evaluate the suitability of alternative options in the current literature (Figure 3; Experimental procedures). Starting from the base year of 2015 (i.e., SDG initiation), progress toward the target for each indicator was measured in percentage terms, according to the standard SDG progress monitoring methodology and terminology,[33] in a range from 0% or less (indicating no or reverse progress; i.e., deteriorating), 0%–50% (i.e., stagnating), 50%–100% (improving), and 100% or greater (indicating that the target has been met or exceeded; i.e., on track).

### Insufficient short-term progress

By 2030, although most SDG targets remained unachievable under the modeled pathways, the individual target achievement varied across pathways, with some resulting in slightly better progress than others but not enough to be on track to fully achieve the SDGs (Figures 4A and 5). To illustrate, we discuss some of the SDGs and pathways by 2030, with improving, stagnating, and deteriorating progress, respectively.

For the 2030 targets, health and wellbeing (SDG 3), quality education (SDG 4), and economic growth (SDG 8) had the highest progress in Green Recovery (82%, 89%, and 97%, respectively; Figure 5) and Fossil-Fueled Development (83%, 89%, and 99%, respectively; Figure S6B). In at least 50% of the realizations for each of these two pathways, progress under SDGs 3, 4, and 8 was either on-track (five targets) or improving (three targets) by 2030 (Figure 4A). A combination of assumptions on human capital investment and low population growth (Figures 2A-I, 2A-V, S1C-I, and S1C-V) put Green Recovery and Fossil-Fueled Development on track toward these targets by 2030. Fragmented World (and then Business As Usual and Inequality) had the slowest progress by 2030, stagnating and even deteriorating from 2015 for most socioeconomic targets under SDGs 3, 4, and 8 (Figures 4A, S4B, and S5B).

Sustainable food (SDG 2) and clean energy (SDG 7) were the two goals with relatively slow progress by 2030 across all pathways. For SDG 2, Fossil-Fueled Development outperformed other pathways by 74% progress, being on track or improving for six of seven 2030 food production and agricultural productivity targets (Figures S6B and 4A). Conversely, progress under Fragmented World was only 36%, being on track in achieving only three food-related targets by 2030 (Figures S4B and 4A). For SDG 7, the progress in Green Recovery is highest (47%), mostly because of economic growth with a higher adoption of efficient end-use technologies and a faster transition to renewable energy (Figure 2C-I). Fossil-Fueled Development and Fragmented World had the slowest progress of 31% and 17%, respectively (Figures S6B and S4B), because of heavy reliance on fossil fuels throughout the century (Figures S1E-V, S1F-V, and S1G-V).

The 2030 progress in biodiversity conservation (SDG 15), responsible production (SDG 11), and climate action (SDG 13) was the lowest in all pathways. By 2030, projected progress toward these targets was either stagnating or deteriorating in all pathways (Figure 4A). Green Recovery aside, this poor environmental performance was largely the result of increasing demand for food production, high meat consumption, and a growing energy-intensive economy in the model input assumptions for these pathways, which posed risks for environmental targets such as agricultural land expansion and nitrogen fertilizer use (Tables S1 and S2). In Green Recovery, despite model assumptions that were expected to counteract environmental damages, the low progress for SDGs 11, 13, and 15 was driven by the momentum of negative trends (e.g., ongoing ecosystem loss, deforestation, and global greenhouse gas emissions) and delayed sustainability improvements from systems change.

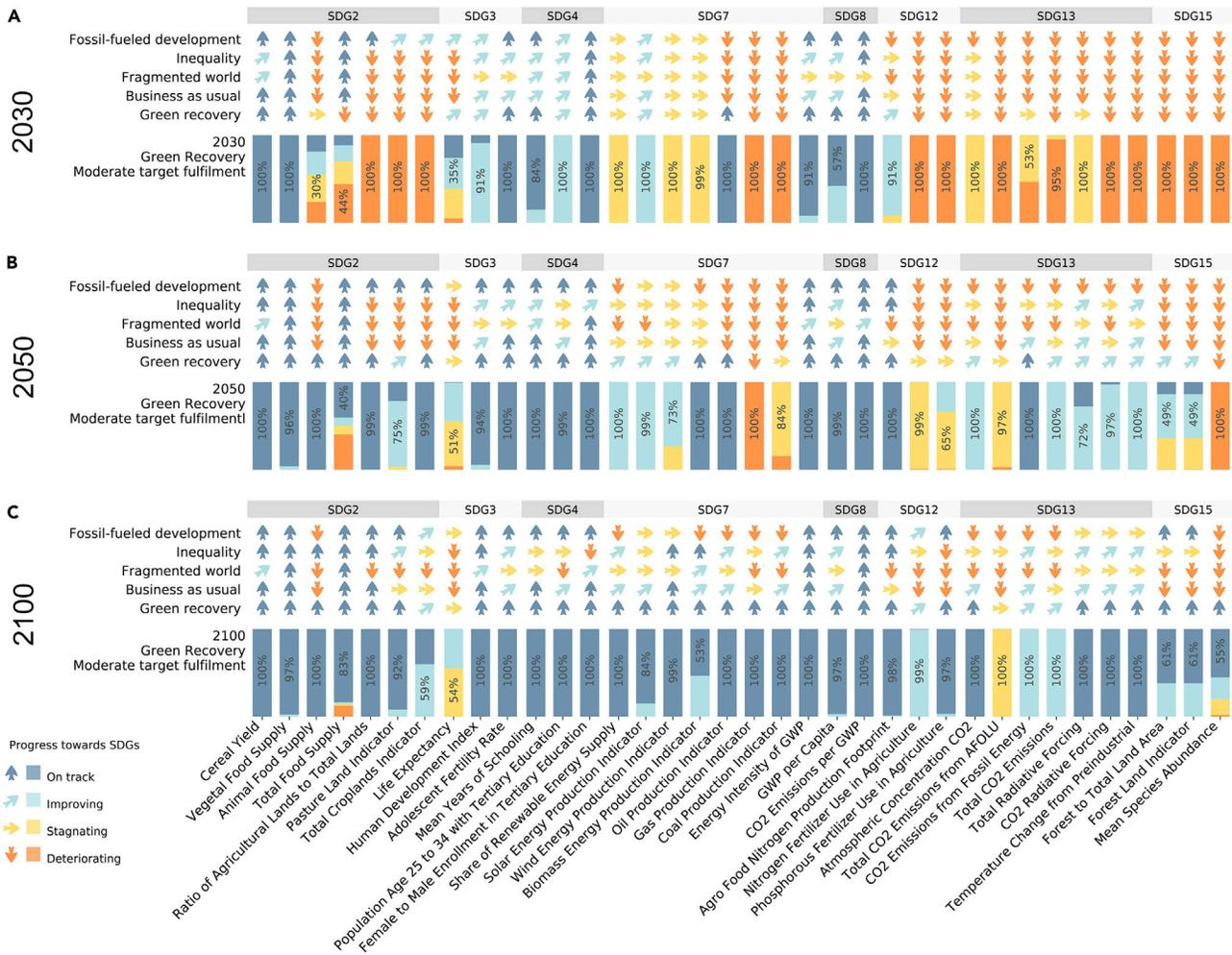

**Figure 4. Projected progress toward moderate SDG targets over time and under five modeled pathways**

(A–C) Progress by 2030 (A), 2050 (B), and 2100 (C). Each column represents one indicator. Related indicators are grouped under SDG headers. Progress levels (deteriorating, stagnating, improving, and on track) at each indicator are color-coded in the stacked bar charts and also represented by arrows for all five pathways (Experimental procedures). The arrows indicate the most likely progress of each pathway across its 10,000 realizations. The stacked bar chart focuses only on Green Recovery as the most sustainable pathway. The annotated percentage in each bar represents the share of 10,000 Green Recovery realizations for the corresponding progress level. For example, in (C), the bar for "Total Food Supply" shows that 83% of the 10,000 possible realizations of Green Recovery had on-track progress, whereas 17% of them had stagnating or deteriorating progress.

### Post-2030 acceleration

Progress towards increasingly ambitious 2050 and 2100 targets accelerated beyond 2030 under all pathways. To illustrate, looking at Green Recovery as the pathway with highest long-term progress, the SDGs that had the worst outcome by 2030 experienced much faster progress toward new targets by 2050 and 2100 (Figures 4B, 4C, and 5).

By 2050 under Green Recovery, progress in responsible production (SDG 12), climate action (SDG 13), and biodiversity (SDG 15) increased to 54%, 74%, and 42%, respectively (Figure 5). Looking out to 2100 with even more ambitious targets than those in 2030 and 2050 (Tables S3 and S4), progress under Green Recovery in these three goals further increased to 94%, 84%, and 90%, respectively (Figure 5). Green Recovery's progress acceleration was less but still significant in other goals as well. For example, progress in food security (SDG 2) and clean energy (SDG 7) reached the highest level among all pathways, to 97% and 99% by 2100, respectively (Figure 5). Even in SDGs where Green Recovery did not seemingly progress much over time (e.g., Health and Well-being; Figure 5), the change in the absolute of value of the related indicators in 2100 is significant (Figures S8B-I to S8B-III). This can be explained by our methodology, which measures the post-2030 progress against shifting targets toward further 2050 and 2100 ambitions and not against the same 2030 target values (Experimental procedures).

Similar acceleration was also observed in Fossil-Fueled Development by 2100, but mostly across socioeconomic goals rather than environmental ones (e.g., SDGs 12, 13, and 15; Figure S6C). Other pathways, such as Fragmented World, also showed non-linear long-term progress, but in the opposite direction, reversing their 2030 achievement and even deteriorating from their 2015

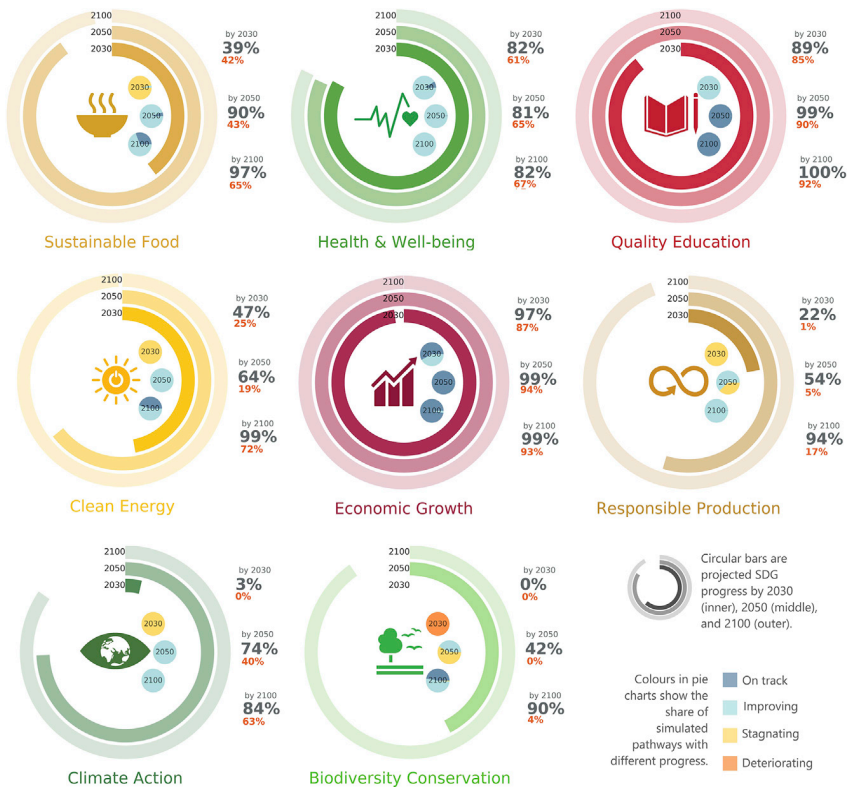

status in socioeconomic and environmental SDGs (e.g., in SDGs 3, 4, 12, 13, and 15; Figure S4C).

The observed acceleration (or deceleration) of progress across pathways is driven by the non-linear systems behavior, leading to time lags and delay between pathway measures and their effects on SDG progress. To illustrate, under Green Recovery, population growth and fossil energy production peaked and then declined around 2050 (Figures 2A-I and 2D-I). Such non-linear behavior underpins the initially slow (by 2030) and later accelerated progress in several SDGs (by 2050 and 2100) that are related to demography and energy systems, such as SDG 7, where lower population and less fossil energy production can directly contribute to its progress.

The non-linear systems behavior characterized by delayed acceleration between pathways and their impacts is driven by a complex chain of system interactions that underlies the SDGs. An example is the initial (i.e., 2030) slow and later (i.e., 2100) accelerated progress in SDG 13 under Green Recovery (Figure 5). The reasons are mixed and manifold. Lower population growth (Figure 2A-I) combined with more sustainable lifestyles can attenuate the increase in energy demand (Figure S1D-I) and long-run impact on energy production, resulting in lower energy sector emissions. In a similar interaction, low population along with exponential growth in access to education over the century (Figure S1C-I) can gradually lead to more environmentally conscious consumption patterns and a higher uptake of healthier and more sustainable diets, as shown by Eker et al.[24] Over time, healthy plant-based diets and lower consumption of high animal-based foods (Figure S1L-I), as the key drivers of land-use and climate change,[34] can reduce the impact of agriculture on land (Figures S1M-I and S1N-I), decelerate or reverse forest loss (Figure 2F-I), and provide significant climate change mitigation (Figure S1O-I).

## Systems change for long-term sustainability

The latest Global Sustainable Development Report suggested different entry points for long-term sustainability.[35] Achieving long-term progress acceleration through these entry points is complex and requires early planning for complementary systems change that cuts across multiple SDGs; changes that should be coherently pursued to transition[36] from currently established to emerging (and more sustainable) socioeconomic and environmental systems (Figure 6).[35] We characterizes systems change for long-term sustainability through the lens of four entry points: (1) human well-being and capabilities, (2) sustainable food systems and healthy nutrition, (3) energy transition and universal access, and (4) sustainable economy decoupled from environmental impacts (Experimental procedures). In each entry point, the scale of change across modeled systems is quantified based on the deviation from the continuation of current reference trajectories (i.e., Business As Usual) to the pathway of highest long-term progress (i.e., Green Recovery) at three timesteps of 2030, 2050, and 2100. With increasing attention to feasibility in modeling studies,[37] we also draw on recent studies to discuss some of the opportunities and challenges on the ground (e.g., new technologies, behavioral change, and grassroots support) and in the broader landscape (e.g., major socioeconomic change, power shift, and policy support) to deepen the understanding of what it would take to facilitate systems change and what could prevent an "idealized" implementation.[36]

### Human well-being and capabilities
Improving education is an essential system change not only for advancing human material health and well-being but also for enhancing human capital in terms of knowledge, skills, and competencies to drive long-term sustainable development.[35] It was at the core of Green Recovery in our modeling as well, reflected in improving access to quality education by 10% (one standard deviation range: 5%–14%) and 40% (23%–53%) compared with 2050 and 2100 Business As Usual trajectories (Figure 7).

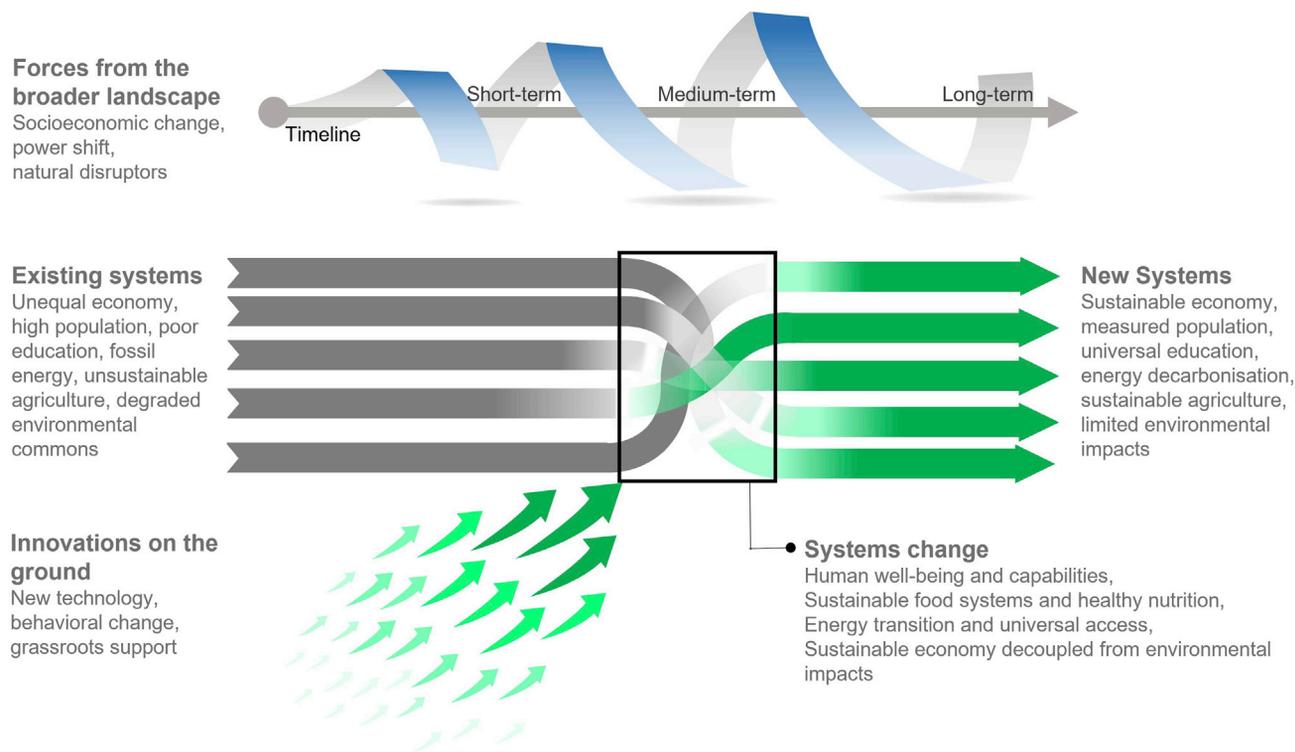

**Figure 6. Conceptualization of long-term sustainability as transformation via complementary systems change**
Systems change can be fostered or impeded by opportunities and challenges on the ground or in the broader landscape (adapted from Geels et al.[36]).

Realizing system change in education cannot be easily achieved in all regions and requires significant technical and political support.[38] Among the support opportunities with proven effectiveness in different contexts are eliminating school fees for universal access to primary and secondary education, improving local access to schools and ending social and legal discrimination to ensure gender equality, and setting up systemic improvement through continuous learning evaluation and enhanced teacher training.[39] The long-term success of these measures rests on overcoming current challenges, such as establishing a stable education system that can allow gradual improvements and shifting mindsets around the role and benefits of inclusive education in less developed regions.[40]

Another important system change to contribute to human well-being and capabilities is related to demography and acting on rapid global population growth.[35] Under Green Recovery, this was represented by reducing population growth by 5% (3%–8%) and 26% (16%–35%) compared with 2050 and 2100 Business As Usual trajectories while improving life expectancy (Figure 7). Improved education with progress in social norms and adoption of bolder actions about family planning with positive impacts on fertility and mortality decline in developing regions are among opportunities for measured population growth.[41] Investment in effective healthcare and newborn health services are other rising opportunities for enhancing prosperity.[42] However, the success of such initiatives may be challenged by the geographic concentration of population growth in emerging economies with a growing middle class (estimated at ~5 billion by 2030) aspiring to lifestyles associated with increased consumption.[43] This highlights the important synergies between lower population growth rates and better redistribution of wealth and how policies addressing inequalities in income and gender could enhance long-term sustainability.[38]

### Sustainable food systems and healthy nutrition
The business-as-usual trajectories and the continuation of current practices for the global food system cannot sustainably and equitably meet the needs of future populations, and the importance of a system change for sustainable food and healthy nutrition is undeniable.[44] One important aspect of this system change is related to land use and limiting agricultural land expansion (which also relates strongly to land as global environmental commons) through more efficient food production with higher yield and productivity. The scale of this change for Green Recovery was 7% (5%–10%) and 10% (7%–13%) reduction in cropland and pasture area, respectively, compared with 2050 and 2100 Business As Usual trajectories (Figure 8) while maintaining sufficient and higher-yield food production (Figures S8A-I and S8A-II). These types of changes can help limit deforestation and reverse biodiversity loss[16] (Figures S8H-II and S8H-III).

Diversified and emerging opportunities exist to control land use change from agricultural activities,[45] such as improvement in crop yields, more efficient use of inputs (e.g., water, nutrients, and pesticides) via automation and precision agriculture,[46] higher livestock productivity (e.g., through better feeding practices and supplements that reduce enteric fermentation),[47]

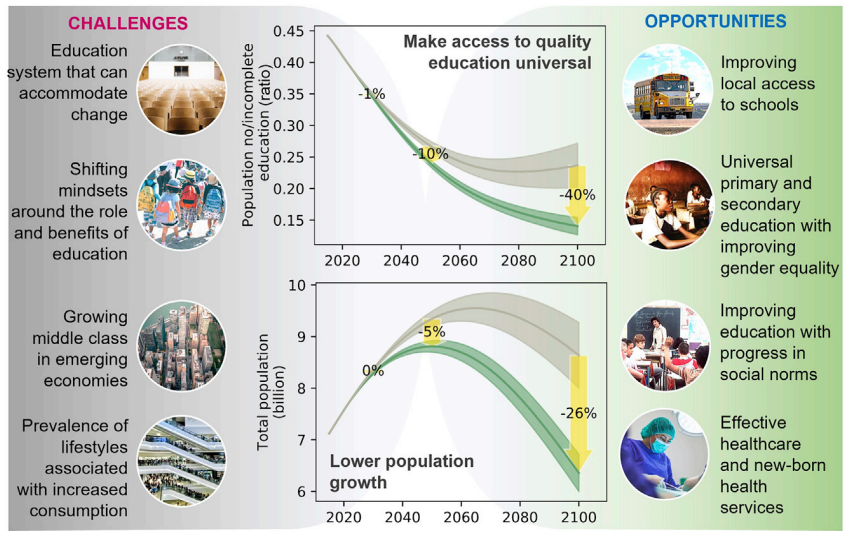

**Figure 7. The underlying systems change in education and demography to shift from Business As Usual to Green Recovery**

In the center plots, envelopes show one standard deviation bandwidth in the results. The middle line is the mean. Yellow arrows show the change percentage needed to deviate from the mean of the Business As Usual envelope to the mean of the Green Recovery envelope in 2030, 2050, and 2100. The mean estimate percentage of improvement (i.e., the distance between the mean value of the two envelopes) is annotated for 2030, 2050, and 2100. Challenges on the left illustrate two examples of potential barriers that can create lock-ins in Business As Usual and impede systems change. Opportunities on the right illustrate two examples of many potential actions that already exist, and their uptake can facilitate systems change.

reducing further demand for agricultural land expansion through controlling food waste via demand-side interventions (e.g., regulations and information/education campaigns),[48] and redesigning agricultural practices (e.g., intercropping and agroforestry).[49] These efforts to limit land use change can, however, face multiple challenges, such as institutional barriers for enabling smallholder farmers to access support and financial resources[50] and the concentration of land ownership in industrial farms, which could be more susceptible and less adaptive to external shocks.[35]

Another important aspect of a sustainable food system is consumption practices and collaborative action on food choices. In Green Recovery, this was translated into 39% (31%–46%) and 50% (43%–57%) reduction in land-based animal (i.e., ruminant meat and dairy) caloric intake in a healthy diet compared with 2050 and 2100 Business As Usual trajectories (Figure 8). More sustainable plant-based diets can improve the health and well-being of communities and also alleviate inequality by helping those affected by the distributional impacts on food supply chains.[51]

Technological innovations, economic incentives, and institutional changes are some of the emerging opportunities to promote healthy diets, among them investment in public health information, guided food choices through incentives, and educational guidelines to promote more nutritious foods.[44,52] Such opportunities, however, rely on significant and rapid behavioral change in the current eating habits of billions of consumers in diverse contexts.[53] This is extremely challenging, given the strong cultural and social norms around diets, such as strong associations between meat and aspects like wealth and masculinity.[54,55] Similarly, the success of many promising technological opportunities, such as novel alternative proteins (e.g., plant-based meats and milks or the prospect of cellular meat or microbial protein) fundamentally relies not only on development of palatable and affordable meat substitutes but also on creating public awareness and normalizing their consumption.[45] Demographic transition to a lower and more educated and prosperous population is among the key enabling factors for such a rapid shift in behavioral and social norms and changing people's attitudes around the potential impacts of their individual choices in the food system (e.g., fewer environmental impacts and lower health risk from less meat consumption).[24]

### Energy transition and universal access

Energy transition is key to economic development and human and social well-being. It can also mitigate current alarming environmental trends, such as increasing emissions and rising temperatures.[17] In Green Recovery, this change was reflected by a decline of 36% (29%–42%) and 80% (75%–84%) in total fossil energy (i.e., coal, oil, and gas) production compared with 2050 and 2100 Business As Usual trajectories, respectively (Figure 9).

There are emerging opportunities that could pave the way for this system change.[56,57] Among them are efforts to increase the share of renewables through a global carbon price scheme with international burden sharing and strong progressive redistribution of revenues to avoid high mitigation costs and trade-offs with poverty,[53,58] financing innovation in renewable energy by private and public financial actors,[59] cheaper renewable energy technologies through subsidies, and a spatially optimized deployment of bioenergy with carbon capture and storage,[60] along with other measures for energy transition in buildings, transportation, and industry sectors.[57] Despite these opportunities, technology and policy and feasibility challenges persist, such as long-term storage of generated renewable electricity and smart grid network management, potential social and environmental trade-offs (e.g., the side effects of biomass and biofuel expansion on land use change), and disproportionate government support (e.g., subsidies) for fossil fuels compared with renewable energy.[35]

Changes in production systems need to be further supported by sustainable consumption practices to ensure reliable, cheap, and clean energy sources. This was reflected in Green Recovery through change in energy consumption patterns to 13% (3%–22%) and 32% (20%–43%) lower energy consumption compared with 2050 and 2100 Business As Usual trends, respectively (Figure 9). A number of key technological and consumption-related challenges need to be overcome to accelerate system change to the pace required. These include a huge stock of older buildings in need of retrofitting of heating and cooling technologies based on renewables, both of which remain out of reach for most consumers in the absence of strong economic incentives.[61] Similarly, transforming energy-intensive, fossil fuel-dependent, and highly polluting industrial processes, increasingly manufactured in developing countries, is challenging too without strong economic incentives and international cooperation and coordination.[62] Key opportunities exist in consumer subsidies to improve dwelling characteristics and incentivize behavior toward larger household sizes with more densely constructed dwellings and energy-efficient appliances.[63] Opportunities also exist to realize low-energy consumption practices through innovations such as digital and artificial intelligence technologies for energy use and monitoring,[64] modern cities with energy-efficient public infrastructure, mobility systems, housing sectors, and smart grid management for long-distance power transmission and less energy loss.[56]

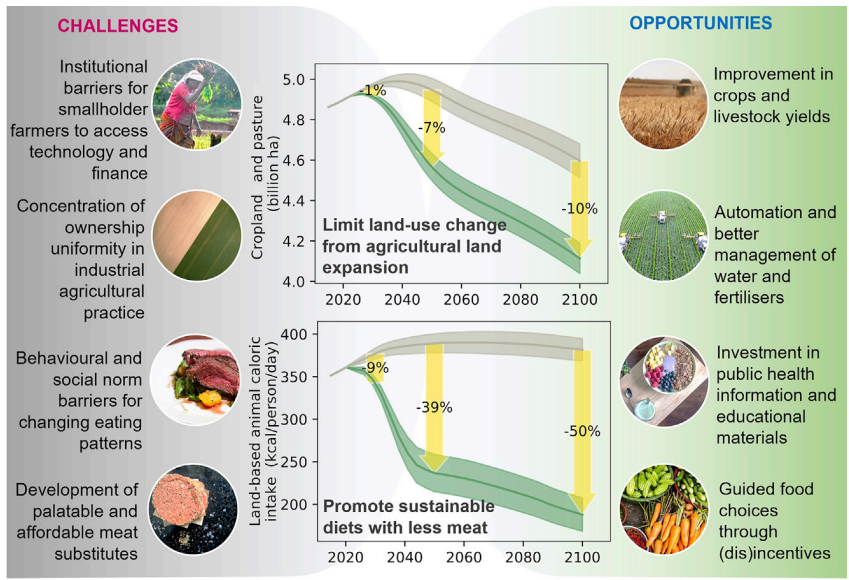

### Sustainable economy decoupled from environmental impacts

A sustainable pathway needs its economic benefits to be decoupled from its environmental costs.[35,65] Advancing human well-being and capabilities, shifting to sustainable food systems and healthy nutrition, and energy transition with universal access together can organically lead to a system change in the broader economy toward sustainable growth with less environmental trade-offs. Under Green Recovery and in terms of economic development, this was represented by sustainable and decarbonized growth of at least 32% (7%–61%) and 52% (5%–118%) higher than Business As Usual by 2050 and 2100, respectively (Figure 10).

Boosting innovation and research can be a key contributor to economic growth. However, this growth can be deeply unequal and therefore unsustainable, resulting in further concentration of wealth and power and environmental exploitation because of overuse of natural resources in less developed regions resulting in a poverty-degradation spiral.[66] Sustainable economic growth needs to also encourage divestment in the current Business As Usual practices and promote innovations that can pave the way for long-term sustainability pathways (green growth) with improved human and environmental benefits.[67] A range of opportunities exists that can support such change, among them support of government science funding mechanisms to guide efforts effectively and with equal opportunities for all,[68] formation of innovation and entrepreneurship incubators to nurture and develop emerging ideas, and support of state investment banks[69] and public-private financing facilities for improved access to financial resources.[70] Among the challenges to realize these opportunities are the immaturity of policies, institutions, and sometimes technologies to promote economies with more efficient use of resources and also the engrained attitudes towards material- and status-related consumption associated with increased wealth.[71]

A sustainable economy with transitioning (food and energy) production and consumption systems can also minimize the environmental impacts, among them the degraded climate system from greenhouse gas emissions, which can have significant impacts on oceanic and terrestrial ecosystem health.[35] Under Green Recovery, the scale of climate change mitigation efforts, represented by the resulting atmospheric $CO_2$ concentration, was 6% (5%–7%) and 20% (18%–21%) lower compared with 2050 and 2100 Business As Usual trajectories, respectively (Figure 10). The climate system is deeply linked to previous systems change, and its emissions mitigation is the result of changes in demography, food, and energy systems. For example, carbon pricing, bioenergy with carbon capture and storage, reforestation, and reduced meat demand are among opportunities from other systems change that can also result in significant impacts and reverse the current climate trends. The emerging support for divestment in polluting industries, increasing green investments, and inclusion of climate change impacts in financial risk management are among important complimentary opportunities to support emissions reduction across all systems.[72] Beyond these, leveraging international governance and global partnerships through currently established frameworks (e.g., the Paris Agreement) and building on emerging public and political will to act on climate change are other opportunities to ensure implementation of systems change in a coordinated manner and effective management of efforts in conflicting contexts.[73]

Opportunities for highly ambitious emissions reduction can be, however, limited by challenges related to their technical feasibility and their significant trade-offs with other systems. For example, faster decarbonization (e.g., 1.5°C pathways[74]), which relies on very high deployment of negative emissions technologies, such as bioenergy with carbon capture and storage, raises concerns with respect to regional availability of geological storage, resource constraints (land or water), and/or securing the social license to support them.[53,75] Even given their feasibility in a context, some of the negative emissions technologies can compete with agricultural production and put food security and biodiversity at risk.[76] Ambitious emissions reduction opportunities therefore need to be further assessed for their policy costs, feasibility, and trade-offs with other non-environmental SDGs before implementation.[37,77]

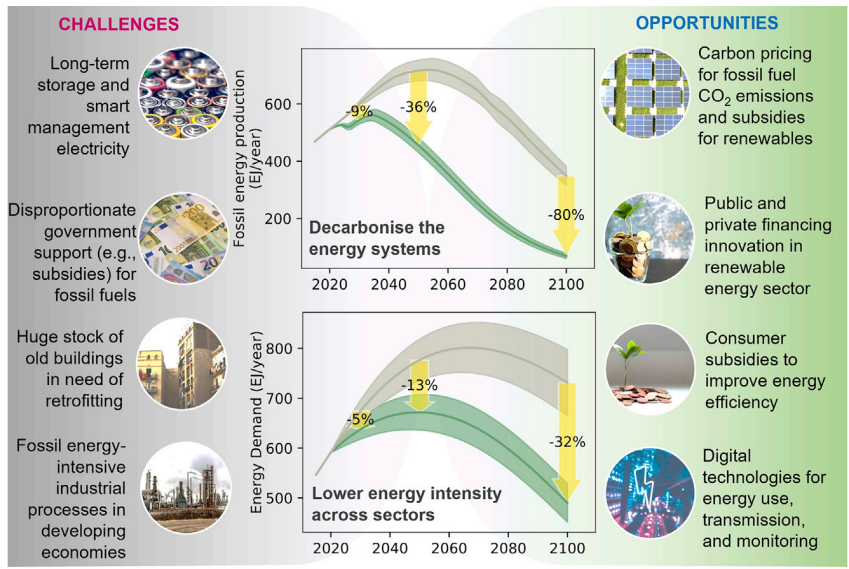

## DISCUSSION AND CONCLUSIONS

### Longer-term assessment in SDG target space

The currently slow progress towards the SDGs poses challenges to the stability of human-natural systems.[78] Calls have been made to revise the SDGs and for new assessments to guide how to lead sustainable development for economic, social, and environmental prosperity.[79] In response, our study aimed to rethink options for the SDGs by adopting a new lens with a century-long timeframe. The longer timeframe (i.e., 2100) allowed simulation of effects of feedback interactions with delayed (post-2030) progress acceleration, which has not been previously discussed in the SDG context. This longer-term analysis is important because it can determine to what extent conditions that appear to make a limited contribution to initial progress by the 2030 milestone could become important later in the century. This long-term perspective can also help plan for the SDGs in an order of priority aligned with future possible trajectories of socioeconomic and environmental development, to better understand potential challenges and opportunities ahead well beyond the 2030 milestone, and to strongly maintain the progress against these challenges in time of diminishing faith and despair.

Any research focusing on a quantitative analysis of longer-term pathways also requires transparent and well-defined formulation of indicators and their desired targets to reveal the gaps and guide actions to fill the gaps.[8] Drawing on recent scientific data and consistent with the 2030 Agenda, our study was novel and complemented recent similar efforts[7,8] in systematically defining a balanced suite of socioeconomic and environmental indicators and setting explicit quantitative targets with increasing ambition levels throughout the 21st century. We used the targets for eval-uating SDG progress over time. We also went beyond that by specifying the critical systems change important for accelerating long-term sustainability. This provided insights into what it would take to shift from Business As Usual to more sustainable pathways over the coming century and what challenges and opportunities could be faced ahead.

This longer-term analysis in SDG target space was, however, limited in some respects and therefore requires future development. First, the longer-term analysis (i.e., to 2100) is challenged by future deep uncertainties in all SDGs, in particular those with potential bigger changes in the future (e.g., related to peace, institution, and implementation). However, our aim was not to predict SDG-specific pathways to the 2100 world with any certainty. Rather, we wanted to constructively use and learn from previous pathway and scenario development, create illustrative pathways for the long-term future, and explore "what if" outcomes of these different pathways for the SDGs. To further advance the treatment of deep uncertainty, future research can use novel scenario discovery techniques[80,81] to obtain more robust insights into future pathways and their long-term impacts on the SDGs. Future research can also examine more systematically the delayed emergence of ambitious pathways to better understand different outcomes for sustainability progress in the medium and long term (e.g., food systems remain Business As Usual over the next few decades, but a major change occurs around mid-century).

Second, our study was also limited by the scope of our model. Although FeliX's global systems were diverse enough to cover most of the key areas of sustainable development[8] pertaining to 36 indicators, they did not cover all systems (e.g., transport, finance, and healthcare), did not span the entire list of 17 SDGs (e.g., those related to well-governed and peaceful societies, which are hard to quantify), or systems change in all possible entry points[35] (e.g., urban and peri-urban development and environmental aspects related to air, soil, and water pollution, which are not included in the FeliX model). Future research can extend the model scope to explicitly represent the missing sectors, better test the implications of sector-specific measures (e.g., subsidies and other incentives to accelerate energy transition), and explore their direct contribution to SDG progress.

### System dynamics modeling for integrated assessment

We used the FeliX system dynamics model to analysis SDG interactions. System dynamics as an established methodology[19] has been used for modeling feedback interactions, delayed response, and non-linear behavior in a climate and sustainability context[23,24,82] and the SDGs in particular.[12,14,83,84] One useful feature of system dynamics models (including FeliX) for sustainability studies is the development of relatively simple and transparent global models where relationships between observed outcomes and modeling assumptions are relatively easy to understand. This is important for SDG analysis in the light of one recent study[85] suggesting that policy-makers are less concerned about accuracy or precision and instead prioritize simplicity and ease of understanding in social and policy processes. Although this relative simplicity could limit predictive power, the link between model accuracy and sophistication remains tenuous.[86,87] Validation of our modeling against past trends[52] and against outputs from more sophisticated models (Figures 2 and S1) also showed a high degree of agreement.

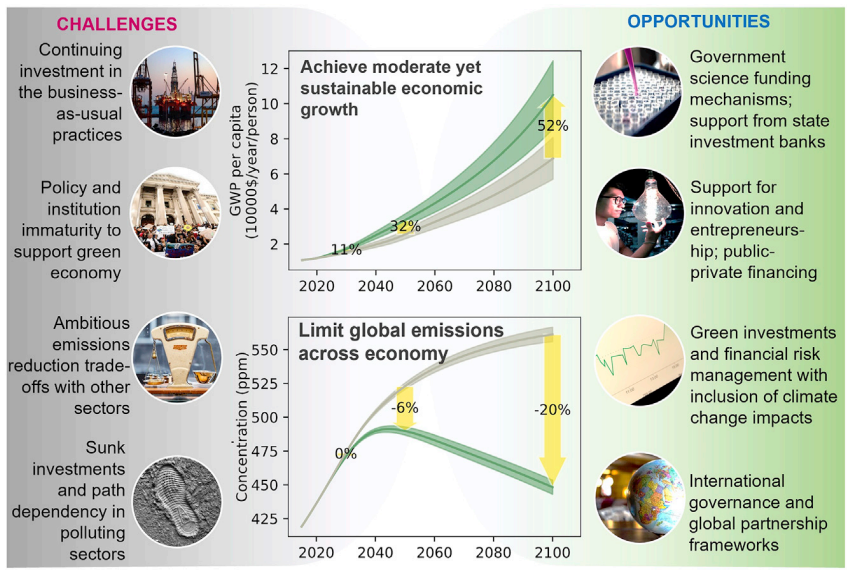

**Figure 10. The underlying systems change in the broader economy and its environmental impacts on emissions reduction in shifting from Business As Usual to Green Recovery**

Another strength of system dynamics models such as FeliX is the endogenous modeling of feedback interactions between social, economic, and environmental systems in one integrated framework (Figure 1; Experimental procedures). These models are ideally suited for capturing non-linearities in short- and long-term pathways, as highlighted in this study. The modeling of feedback interactions can help enhance the understanding of SDG inter-connections and complexities (e.g., non-linearity, tipping points, and delays).[12] Understanding complex interactions can lead to insights around SDG synergies and trade-offs by identifying underlying mechanisms of barriers or policy resistance and designing synergistic solutions that can translate to a more successful outcome for sustainable development.[88] Aggregate and descriptive system dynamics models can be used to complement the insights from other integrated assessment models (e.g., Earth systems and partial or general equilibrium) that focus more on a detailed view of biophysical and socioeconomic systems than feedback between systems.[89,90]

Although FeliX represents several key system elements and their feedback, it does not capture all important interactions. Future research can contribute to this by identifying and incorporating other feedback interactions currently not represented in existing integrated assessment models, including in FeliX. Examples can include modeling climate feedback interactions with other systems that are important in the context of sustainability, such as agriculture (e.g., $CO_2$ fertilization effects on natu-ral vegetation and crop growth), land use (e.g., prolonged precipitation effects on land management decisions), energy (e.g., rising temperature effects on energy demand), and human behavior (perceived climate extreme event risks alter human emissions).[89,90]

### Co-designing pathways for local contexts

Our modeling was focused at the global level. Global-scale studies[12,23,24] are important for their role in capturing interactions between systems, monitoring their aggregate outcomes, and guiding harmonized high-level interventions to ensure universal progress toward the 2030 Agenda and beyond.[91] Several of the environmental challenges addressed are common worldwide (e.g., temperature increase and biogeochemical flows), and management relies on an understanding of their aggregate effects and the globally connected systems that underpin them. Despite this global connection, different locations face unique, place-specific issues and have their own needs and sustainability priorities, creating strong geospatial ties for many of the SDGs.[92] For example, although unsustainable diets are a common challenge with an impact on global emissions shared by all, their effects on food demand, food production, and land use change vary between locations and depend on the unique socioeconomic and environmental characteristics of each region, including social norms, education level, resources, and dominant food systems of each community.[53] This necessitates future research to translate and downscale the global understanding of pathways and the SDGs at the local level to better acknowledge indigenous values, cultural differences, available resources and technologies, and local political and governance frameworks and also better understand the distributional effects and variations of progress across scales.[15]

However, there are often significant limits in modeling and translating the implications at the local scale, driven by the challenges of understanding heterogeneities on the ground[93] and resolving fundamental disagreements among stakeholders.[91] Previous studies have suggested frameworks for addressing these challenges through transdisciplinary approaches that can go beyond working with researchers and facilitate public community engagement in pathway development processes.[94,95] Although still a niche field, the growing application[96,97] of a variety of transdisciplinary approaches, such as knowledge co-production (including local, practical, and indigenous knowledge) and participatory processes with stakeholders (i.e., co-designing pathways, local priorities,[10] and plans[98]) have provided opportunities to advance the local-scale understanding of the SDGs.

## Pathway development for bolder sustainability actions

Despite originally having been developed for climate projections,[26] the SSP-RCP compliant pathways and their variations have been widely used as benchmarks in broader sustainability science[5,7,14] and its related areas (e.g., water,[99] agriculture,[100] and biodiversity[101]). Their adaption in the current study was partly motivated by their extensive coverage of global-scale socioeconomic and environmental parameters needed for SDG analysis. SSP-RCP-compliant pathways also provided a basis for generating alternative plausible futures with fundamentally different sustainability outcomes, suitable for this study, where the emphasis was to demonstrate the range of alternative futures (spanning from low progress) to the end of the century.

However, the five selected pathways in the current study were only used as illustrative archetypes to highlight specific variation across most commonly used combinations of SSP-RCP pathways[26] and were not intended to cover all future possibilities. For example, we did not include pathways of higher climate change mitigation[17] because of the trade-offs with other SDGs. Pathways from the SSP-RCP frameworks also do not include explicit driving forces related to some of the SDGs, such as gender inequality or partnerships that may impact projections because of missed synergies and trade-offs with other goals in FeliX. These issues would benefit from future research that goes beyond the current ambitions in SSP-RCP frameworks in building SDG-specific pathways for bolder actions not only for the 2030 Agenda[7] but also for longer-term timeframes (i.e., 2100) that can show acceleration toward sustainability later in the century.[53]

## EXPERIMENTAL PROCEDURES

### Resource availability

#### Lead contact
Further information and requests for resources and reagents should be directed to and will be fulfilled by E.A.M. (e.moallemi@deakin.edu.au)

#### Materials availability
All new materials are provided via links under "Data and code availability."

#### Data and code availability
The full code, results, and datasets used and generated are available at Zenodo: https://doi.org/10.5281/zenodo.6459874. Tables S2 and S4 in the supplemental information can be accessed at Zenodo: https://doi.org/10.5281/zenodo.6609917. FeliX, the simulation model used in this study, is available at Zenodo: https://doi.org/10.5281/zenodo.6459874 and from the IAMC website: https://www.iamconsortium.org/resources/model-resources/felix/.

### The FeliX system dynamics model

FeliX is a system dynamics model that simulates complex interactions among 10 global systems: population, education, economy, energy, water, land, food (including diet change), carbon cycle, climate, and biodiversity. FeliX was originally developed for projecting socio-environmental impacts in human-natural systems[23] and was later advanced for exploring emissions pathways[23,102] and evaluating sustainable diet shift.[24] FeliX is one of the very few models of human-natural systems that covers the breadth of social, economic, and environmental aspects (and their feedback interactions) in one integrated framework suitable for SDG analysis. The model operates at an annual timescale and is designed to project global-scale future socioeconomic development and environmental conditions over the long term to 2100. It is implemented in the Vensim software and has been calibrated with historical data from 1900–2015 (see Rydzak et al.[22] for calibration results and graphs).

A key validation method in system dynamics and other modeling methodologies that project future pathways is based on comparing them with historical data (to show whether this model is at least reliable for reproducing the past) and with the future projection of other models, if available (to show the new model produces sensible results, also called cross-validation). FeliX has been validated in both ways. For validation with historical data, refer to the extended technical report for FeliX,[22] which includes detailed validation of each of FeliX's sub-models against historical data. For validation against other models, see Global systems modeling and Figures 2 and S1.

The use of system dynamics as a methodology has a long history. One of the first and most enduring applications of system dynamics was in The Limits to Growth's modeling of the environmental and social impacts of global industrialization in 1972,[103] pointing out that ecological and economic stability would not be out of reach if actions were taken early.[104] Since then, system dynamics has been used widely as an established methodology in sustainability science (for a review, see Moallemi et al.[83] and Allen et al.[105]). System dynamics models can be (and are often) developed based on a co-design process that enables interaction between researchers and stakeholders and supports synthesis of disciplinary and transdisciplinary knowledge.[106] This and other features mentioned in the main text were also highlighted in the review by Allen et al.[105] of modeling tools for the SDGs, suggesting that system dynamics models can be more transparent and legitimate compared with other modeling approaches.

Despite the methodological advantage, FeliX misses some sectors and requires future improvements. For instance, the primary energy demand in transport (~15% of global GHG emissions in 2019[53]) is expected to increase by 25% by 2050 in a Business As Usual (BAU) scenario,[107] contributing to an increase of around 25% in global primary energy demand,[108] but an explicit modeling of the transport sector is missing in FeliX. Although our uncertainty exploration of energy demand projections (17%–33% increase in global primary energy demand by 2050 and compared with 2020 in Business As Usual; Figure 5) could cover the implications of transport system indirectly, a future improvement to model the transport sector (along with other missing sectors, such as governance) endogenously would be needed for better projections.

A summary of the sectoral modules in FeliX is available in Figure 1, and a detailed description is provided below. Important interactions among the eight SDGs modeled in FeliX are available in Table S5. Readers are also referred to the original FeliX documentation[22] and previous papers that have used FeliX[23,24] for an extended description and validation of the model with respect to historical data and cross-comparison with other scenarios. The model and its supporting data are publicly available online (Data and code availability). The equations underlying each SDG indicator in the model are available in the supplemental information. Information about the methodology for computing indicator values in the model is available in the supplemental experimental procedures.

### Population and education

The population module describes population growth based on an aging chain and computes the male and female population size of 5-year age cohorts between the ages of 0 and 100+. The birth rate, driven by education and gross domestic product (GDP) per capita, is the main factor affecting population dynamics (either growth or decline), alongside the reproductive female population represented by gender and age-cohort segmentation in the model. The chain structure in the model represents the transition of newborns through the age cohorts as they age, meaning that each age cohort except the "0–5" cohort has one inflow (maturation of the previous cohort) and two outflows (maturation to the next cohort and mortality rate). In the population sector, gender differences are taken into account in two respects: the gender fraction of newborns, representing female infanticide, and educational enrollment and graduation differences. The population module also computes change in life expectancy with impacts for health services, food, and climate risk. Population is the core module in FeliX impacting, directly or indirectly, all other sectors, such as energy demand, water use, effects on fertilizer use, and food consumption. The population size at different age cohorts feeds into the education module to compute the population of primary, secondary, and tertiary education graduates through the feedback loops among the enrollment rate, graduation rate, and persistence to eventually reach the last grade of each education level. The accumulation of the educated population in all age cohorts between 15 and 64, multiplied by a labor force participation fraction, computes the labor force input for the economy module. Population and education are calibrated with the historical demographic data from the United Nations (UN) Department of Economic and Social Affairs.[109]

### Economy
The economy module is modeled as a Cobb-Douglas production function, where total Gross World Product (GWP) is computed from labor input, total capital input from the energy and non-energy sectors, and total factor productivity from energy and non-energy technologies. FeliX further develops the Cobb-Douglas function to incorporate the impacts of changes in ecosystems and climate change on the economic outputs. Given that human development should include measures beyond economic advances, FeliX also computes an alternative measure, called human development index, which is an indicator of health (life expectancy), educational attainment, and income. The economy module is calibrated with historical statistics of world economy[110] and United Nations Development Programme (UNDP) data.[111]

### Energy
The energy module models energy demand as a function of GDP per capita and population. Energy consumption is modeled through the market share of different energy sources by capturing the price-competitive mechanisms between three fossil (coal, oil, and gas) and three renewable (solar, wind, and biomass) sources. Energy production from each fossil source is modeled as a function of energy demand, the market share of energy source, the effect of investment on energy production, and the identified fossil energy resource. FeliX models the technological advancement in discovery of fossil resources and investment in exploration to account for undiscovered resources that can be identified in the future. FeliX also models the technological improvement for recovery of fossil resources. The basic model structure for renewable energy sources is similar to fossil fuels, determined by five key submodules of available renewable resources (e.g., average sun radiation and wind available area), the supply chain of installed capacity and their aging process, the unit cost of production (e.g., the impact of wind and solar technology learning curve), available investment, and technological efficiency and productivity (e.g., solar conversion efficiency and wind capacity factor). The energy module is calibrated with data from the International Energy Agency (IEA).[112]

### Water
FeliX models the water sector through water scarcity; that is, the balance between water supply and withdrawal. Water supply is a function of available water resources, a drought rate for the impact of climate change, water withdrawal from different sectors, and the recovery of water used in those sectors. Water withdrawal is for agriculture, industrial, and domestic sectors. Agricultural water withdrawal depends on irrigated and rainfed agricultural lands, industrial water withdrawal depends on GWP (economic activities), and domestic water withdrawal depends on population and GWP. See The water module is calibrated with historical data from Intergovernmental Hydrological Programme (IHP), The United Nations Educational, Scientific and Cultural Organization (UNESCO).[113]

### Land
The land sector in FeliX is distributed among four categories of land use: agricultural, forest, urban/industrial, and "other." Land use can be repurposed and switch between types depending on demand for more agricultural land. Demand for agricultural land is balanced by increasing crop yields via fertilization. Agricultural land is divided into arable land, permanent crops, and permanent meadows and pastures. Arable land and permanent crops can be harvested to produce food and feed as well as energy crops for biomass. Permanent meadows and pastures can only be used for food production. The area of arable lands harvested is driven directly by food, feed, and energy crop production and indirectly through food demand and biomass energy demand. Crop and livestock yields are modeled as a function of input-neutral technological advancement, land management practices (impact of economy), water availability (impact of drought), nitrogen and phosphorus fertilizer use, and climate change (impact of carbon concentration). Nitrogen and phosphorus fertilizer use in agriculture, from commercial sources or produced with manure by pasture- and crop-based animals, is explicitly modeled in FeliX. However, potash fertilizer is not included because it constitutes the smallest fraction of global fertilizer use (~20%), and its environmental impacts are much lower compared with nitrogen and phosphorus because of high efficiency of uptake and low leakage rates.[114] Change in forest land cover is modeled through conversion with other land uses as well as harvested forest areas needed for biomass energy production. Forest land fertility is modeled endogenously as a function of the effect of biodiversity, land management practices, climate change, and $CO_2$ concentration. The land module in FeliX is calibrated with global scale historical data from Food and Agriculture Organization Statistics (FAOSTAT).[115]

### Food and diet change
The food module in FeliX includes food demand and supply (including waste fraction) as well as diet shift in food consumption of the population. Food demand is a function of food and feed fraction in demand, each of which is determined based on the size of the population with animal-based and vegetable-based food diets. Food supply is the sum of the supply of animal-based products, including crop-based meat (poultry and pork), pasture-based meat (beef, sheep, and goat), dairy, eggs, and the supply of plant-based products, including grains, pulses, oil crops, vegetables, roots, and fruits. Food production (related to food supply) depends on the area of harvested lands (from agricultural lands) and the crop and livestock yields (already discussed in the land module). The food consumption (related to food demand) is determined by linking to a model that relates human behavior and dietary choices to different population segments (e.g., male and female, level of education). The diet change model[24] explains various environmental actions to move toward more sustainable (less meat) diets based on two feedback mechanisms: diet change because of social norms and diet change because of a threat and coping appraisal. The latter is linked to threats from climate events as an important feedback structure between physical and human systems. The food and diet change module is calibrated with historical data from FAOSTAT and Global Burden of Disease datasets.

### Carbon cycle
FeliX models $CO_2$ emissions endogenously based on the accumulation of carbon emissions from the energy and land sectors in the atmosphere. $CO_2$ emissions from land include emissions from agricultural activities (i.e., food production and land use change to agricultural lands) as well as deforestation and forest conversion to managed forests and plantations. $CO_2$ emissions from the energy sector are computed explicitly based on the carbon intensity of energy production from fossil and renewable sources. Emissions from the energy sector also capture endogenously the effect of improvement in carbon capture and storage technology and a desired emissions level from fossil fuels. Carbon is cycled through terrestrial reservoirs, gradually absorbing into the biosphere, pedosphere, or oceans based on C-ROADS,[116] a climate model also used for climate impact analysis by The United Nations Framework Convention on Climate Change (UNFCCC). Carbon dissolution into the ocean is through the mixed ocean layer (depth, 0–100 m) and subsequently through four modeled deeper layers (100–400, 400–700, 700–2,000, and 2,000–2,800 m). See Walsh et al.[23] for modeled equations of carbon flux among different reservoirs. The carbon cycle module is calibrated with historical emissions data from the Carbon Dioxide Information Analysis Center.[117]

### Climate
The climate module models $CO_2$ radiative forcing endogenously based on accumulated carbon (from land and energy) in the atmosphere compared with the pre-industrial level. Radiative forcing of other gases ($CH_4$, $N_2O$, and HFC) is modeled by linking FeliX to RCP scenarios and reading data from the projected forcing levels with the marker models of the SSPs (i.e., IMAGE, GCAM, AIM, and MESSAGE). The effect of total radiative forcing is associated with temperature anomalies as in the C-ROADS model. The surface temperature change is also affected by negative (cooling) feedback because of outbound longwave radiation as well as heat transfer from the atmosphere and mixed ocean layer to the four deep ocean layers.

### Biodiversity
FeliX captures the effect of changes in land cover, land use, and climate impact on the species carrying capacity (global average). The biodiversity module uses this carrying capacity to compute the mean species abundance from the species regeneration and extinction rates. The biodiversity module was calibrated with historical data from the Secretariat of the Convention for Biological Diversity database.[118]

### Pathway simulation
A complementary set of socioeconomic and environmental assumptions was identified from the current pathway projection literature to be used as FeliX inputs for future pathway projections. These assumptions were informed by the SSPs and the RCPs as widely used scientific frameworks for capturing a range of long-term uncertainties with a manageable number of alternative futures.[26] These frameworks have also been used frequently in several previous sustainability assessments.[7,14,100,101]

Among various SSP-RCP combinations, we selected five benchmark pathways of SSP1-RCP2.6, SSP2-RCP4.5, SSP3-RCP7.0, SSP4-RCP6.0, and SSP5-RCP8.5 for projection with the FeliX model. The pathway assumption space included the global trends of different socioeconomic and environmental driving forces to 2100. They spanned socioeconomic (fertility, mortality, migration, educational attainment, and economic growth), energy and climate (energy demand, technology advances, fossil resource extraction, and production cost), land (land use change, crop and livestock yields, and land productivity), food and diet (waste, consumption, and diet change), emissions trajectories (1.9, 2.6, to 4.5, to 6.0, and 8.5 W m$^{-2}$ of global radiative forcing to 2100), and their associated climate policies (Table S1). The defined pathway assumption space was translated into relevant quantitative values for the FeliX's parameter settings (Table S2) using Vensim's built-in function (i.e., Powell) which is often used for quantifying system dynamics model parameters.

FeliX has many parameters, and therefore an evaluation of the impacts of uncertainty in parametric assumptions is necessary. To evaluate the effects of uncertainty, a global sensitivity analysis was performed to identify influential parameters whose uncertainty could have important impacts on pathway projections. Among the global sensitivity analysis methods, Morris elementary effects is ideal for integrated assessment models that have a large number of input parameters and require generation of reliable results with high computational efficiency[119] (Figure S7A). When the influential parameters were identified, to understand the full scale of variation in pathway performance in response to these uncertainties, a series of model runs was conducted using Latin hypercube sampling. Each run is a computational experiment, showing a realization of each pathway. We simulated 10,000 runs (realizations) of each pathway (50,000 total across all pathways).

The resulting projections and their uncertainty range were compared across socioeconomic and environmental output variables with the projections of other models, including IMAGE, MESSAGE-GLOBIOM, AIM, GCAM, and REMIND-MAGPIE[29] to assess the level of (dis)agreement with other models in pathway projections (Figures 2 and S1).

### SDG progress measurement

The SDG framework includes 17 goals and 231 indicators to measure progress towards 169 targets, but they are too broad and complex to support quantitative assessment.[8] Therefore, we operationalized the SDGs in FeliX by selecting a subset of indicators, setting science-based targets for the selected indicators, and measuring progress toward targets as below.

#### Indicator selection

A list of 36 SDG-related indicators was selected from the United Nations Statistical Commission (UNSC) and other sources (e.g., Organisation for Economic Cooperation and Development [OECD], World Health Organization [WHO], United Nations Food and Agriculture Organization [FAO], and World Bank) based on three criteria (Figure 3). First, we looked at the global relevance of the potential indicators for measuring SDG progress (SDG applicability). Second, we assessed the ability of FeliX to quantify the SDG indicator (model fidelity). For indicators that were not present in FeliX, we either advanced the model structurally or chose proxies (i.e., a variable that is closest to the SDG indicator). For example, we did not include an official indicator for biodiversity conservation, such as the red list index, because the required data are not produced in FeliX. Instead, we presented mean species abundance as a proxy indicator for biodiversity.[16] Third, we ensured that the selected indicators are amenable to specification of quantitative performance thresholds for measuring progress towards the SDGs (target relevancy). All indicators that passed these three criteria were included in the analysis.

#### Target setting

Successful evaluation of progress towards the SDGs required a science-driven characterization of targets as quantitative thresholds on each indicator. We defined targets for each indicator using a four-step decision tree (Figure 3). First, we used available quantitative thresholds that were explicitly reflected in the official SDG framework to set targets (SDG absolute threshold; 3 indicators). For example, SDG 8 indicates "at least 7 per cent GDP growth," which can translate into a specific target for the growth rate of the "GDP per capita" indicator.

Second, if an explicit target was not mentioned in the SDG framework, then we used a technical optimum to set targets (technical optimum; 27 indicators). We used targets, wherever relevant, that were identified in other scientific journal articles, global reports,[33,120] and online databases.[121] For example, we used the IPCC's levels of radiative forcing for keeping the global temperature below 1.5°C as target levels for the "radiative forcing" indicator.

Third, wherever the SDG absolute threshold and technical optimum were not applicable, we followed the 2030 Agenda's principle of "leave no one behind" and set the targets based on the average state of the top performing countries in a base year using historical documented data (leave no one behind; 5 indicators). Here the global average as calculated by FeliX is expected to reach the levels of current top performing countries. In selecting the top performing countries, we removed the outliers from the list to reduce bias in our calculation. For example, a small country with limited arable land typically has very low levels of fertilizer application. Therefore, inclusion of this country as a top performer in calculating the target for the "food and agriculture phosphorous balance" indicator can be misleading for larger countries with a larger contribution to global food production. Where performance data were not available at the country level, we used regional data (e.g., OECD and continents).

Fourth, in the absence of any relevant targets, we nominally set a sensible improvement target in the indicator value from the world average in a base year guided by historical data (sensible improvement; 1 indicator). For example, "total $CO_2$ emissions from agriculture" is an indicator with no absolute threshold mentioned in the original SDGs or technical optimum in other studies. The value of this indicator is also sensitive to the size of a country's agricultural sector. Therefore, leaving no one behind and the average of the top performers did not lead to a meaningful target. In this case, we used a level of global improvement as a target for the indicator.

For each indicator, three target levels were set for selected indicators (weak, moderate, and ambitious) to acknowledge different levels of ambition in target setting and the high sensitivity of pathway performance to target specification. At each level, targets were set for 2030, 2050, and 2100 in alignment with the major global sustainability milestones. All results in the main text are based the moderate targets. The results for ambitious and weak targets are available in the supplemental information. The target values and their justification are available in Tables S3 and S4.

#### Progress quantification

To measure progress toward targets at each indicator, we normalized indicator values, each of which had different scales and units of measurement, to ensure comparability and consistent interpretation. For each target, we normalized indicator values to represent performance against target achievement, ranging between the 0% (no progress or divergence away from targets) and 100% (meeting or exceeding targets). The higher values denote a better performance, and the gap from 100 indicates the distance that needs to be taken to achieve the target. The scores below 0 and above 100 were interpreted as where the world is deteriorating from the status quo and exceeding target levels, respectively. The indicator values were normalized based on the rescaling formula in Equation 1,

$$I_{ij}(x_i, w_i, t_i) = \frac{x_i - w_i}{t_i - w_i} \times 100 \qquad \text{(Equation 1)}$$

where $I_{ij}$ is the computed normalized value of indicator $i$ under goal $j$, $x_i$ is the model estimate of indicator $i$ in a single projection, $w_i$ is the base year (FeliX) value in 2015, and $t_i$ is the indicator target level for a certain year. We then aggregated the normalized indicator values into an index score to represent global progress toward each SDG (Equation 2),

$$I_j'(N_j, I_{ij}) = \sum_{i=1}^{N_j} \frac{I_{ij}}{N_j} \qquad \text{(Equation 2)}$$

where $I_j'$ is the SDG and $N_j$ is the number of modeled indicators under goal $j$. The index and its methodology were adopted from a similar index used in global monitoring of the SDG progress.[33] We used the arithmetic mean with a normative assumption of equal weight across each goal's indicators to align with the global efforts to treat all indicators equally and only prioritise indicators when progress is lagging. This also assumes that there is unlikely to be a consensus on SDG indicator priorities. Based on the normalized values at the indicator level and aggregated indices at the goal level, we measured world progress toward targets at four levels. "On track" indicates progress highly likely to achieve (or exceed) global sustainability targets (i.e., indicator and goal level target achievement $\geq$ 100%). "Improving" indicates positive trends toward the goal and indicator level targets but meeting them is unlikely, so challenges remain (i.e., target achievement between 50% and 100%). "Stagnating" indicates performance following current trends, little chance of target achievement, and significant challenges remain (i.e., target achievement between 0% and 50%). "Deteriorating" indicates a reversing trend (i.e., target achievement $\leq$ 0%).

### Systems change characterization

We characterized the nature and scale of systems change required to ensure that the pre-conditions are in place for long-term SDG progress and discussed their challenges and opportunities ahead. We specified systems change in relation to four of the entry points that were within our model scope, originally discussed in the Global Sustainable Development Report:[36] (1) human well-being and capabilities, (2) sustainable food systems and healthy nutrition, (3) energy transition and universal access, and (4) sustainable economy decoupled from environmental impacts. To characterize systems change in each entry point, we first selected one variable from our model outputs that could best represent each system and its associated entry point. They included total population (billion) and population with no/incomplete education (ratio) for the first entry point, cropland and pasture area (billion ha) and land-based animal caloric intake (kcal person$^{-1}$ day$^{-1}$) for the second entry point, energy demand (EJ year$^{-1}$) and fossil energy production (EJ year$^{-1}$) for the third entry point, and GWP per capita ($10,000 person$^{-1}$ year$^{-1}$) and atmospheric $CO_2$ emissions (ppm) for the fourth entry point.

Second, we measured the scale of change in each selected output variable based on the distance between a reference pathway and Green Recovery in 2030, 2050, and 2100. Given future uncertainties, we measured a range including the mean and one standard deviation of this distance between the two pathways. It is worth noting that, across all output variables (i.e., systems change), depending on what the reference pathway is, the scale of change required to shift to Green Recovery can vary. The quantified scale of change here is based on deviation from the Business As Usual pathway (SSP2-4.5), whereas assuming other pathways as a reference (e.g., SSP3 and SSP5) can lead to much larger deviation.

To identify the drivers of systems change, we first identified high-impact model parameters that can drive change in population, education, economy, land, food, energy, and climate systems based on FeliX's sensitivity analysis results (as discussed for pathway projection and shown in Figure S7A). The goal was to find the combinations of high-impact parameters that can be most predictive of systems change. Those high-impact combinations (Figures S7A) were categorized according to influence in relation to the systems change under each entry point (Figure S7B). For each system change and in relation to its drivers, we discussed some of the challenges and opportunities qualitatively based on what has been identified previously in other studies. The goal was to enable a deeper understanding of the feasibility of our modeling.

### SUPPLEMENTAL INFORMATION

Supplemental information can be found online at https://doi.org/10.1016/j.oneear.2022.06.003.

### ACKNOWLEDGMENTS

B.A.B. and E.A.M. received funding from The Ian Potter Foundation and Deakin University for this research. E.A.M. also acknowledges funding from the 4TU.HTSF DeSIRE program of the four universities of technology in the Netherlands. The graphical abstract was designed by Ooid Scientific. Images in Figures 7, 8, 9, and 10 are by Denisse Leon, Doug Linstedt, Nathan Dumlao, Note Thanun, Artur Tumasjan, Tuan Nguyen, Heamosoo Kim, Louis Reed, Ying Ge, Maria Lupan, Katie Rodriguez, meriç tuna, Erwan Hesry, Nandhu Kumar, Jules Bss, Jannis Brandt, Megan Thomas, Jason Leung, LikeMeat, Anders J, micheile, Roberto Sorin, Ibrahim Boran, Max Bender, Jessica Pamp, Patrick

Hendry, Markus Winkler, Kyle Glenn, Piret Ilver, and George Evans, all on Unsplash under free license.

### AUTHOR CONTRIBUTIONS

Conceptualization, E.A.M.; methodology, E.A.M., S.E., and L.G.; investigation, E.A.M., S.E., L.G., B.A.B., M.H., J.K., P.M.R., M.O., and Z.G.; visualization, E.A.M.; writing – original draft, E.A.M., B.A.B., S.E., and M.H.; writing – review & editing, E.A.M., B.A.B., S.E., L.G., M.H., Q.L., J.K., P.M.R., M.O., and Z.G.

### DECLARATION OF INTERESTS

The authors declare no competing interests.

Received: August 11, 2021
Revised: April 18, 2022
Accepted: June 14, 2022
Published: July 7, 2022